\documentclass[prl,preprint,superscriptaddress,floatfix,showpacs]{revtex4-1}

\usepackage{amsmath,amsfonts,amssymb}
\usepackage{graphicx,epstopdf}
\usepackage{tabularx}
\usepackage{bbold}
\usepackage{bm}
\def\beq{\begin{equation}}
\def\eeq{\end{equation}}
\def\beqn{\begin{eqnarray}}
\def\eeqn{\end{eqnarray}}

\begin{document}

\title{Enhanced many-body effects in the excitation spectrum of a weakly-interacting rotating Bose-Einstein condensate}

\author{Raphael Beinke}
\affiliation{Theoretische Chemie, Physikalisch-Chemisches Institut, Universit\"at Heidelberg, Im Neuenheimer Feld 229, D-69120 Heidelberg, Germany}

\author{Lorenz S. Cederbaum}
\affiliation{Theoretische Chemie, Physikalisch-Chemisches Institut, Universit\"at Heidelberg, Im Neuenheimer Feld 229, D-69120 Heidelberg, Germany}

\author{Ofir E. Alon}
\affiliation{Department of Mathematics, University of Haifa, Haifa 3498838, Israel}
\affiliation{Haifa Research Center for Theoretical Physics and Astrophysics, University of Haifa, Haifa 3498838, Israel}

\date{\today}

\begin{abstract}
The excitation spectrum of a highly-condensed two-dimensional trapped Bose-Einstein condensate (BEC) is investigated within the rotating frame of reference. The rotation is used to
transfer high-lying excited states to the low-energy spectrum of the BEC. We employ many-body linear-response theory and show that, once the rotation leads to a quantized vortex in the ground state, already the low-energy part of the excitation spectrum shows substantial many-body effects beyond the realm of mean-field theory. We demonstrate numerically that the many-body effects grow with the vorticity of the ground state, meaning that the rotation enhances them even for very weak repulsion. Furthermore, we explore the impact of the number of bosons $N$ in the condensate on a low-lying single-particle excitation, which is describable within mean-field theory. Our analysis shows deviations between the many-body and mean-field results which clearly persist when $N$ is increased up to the experimentally relevant regime, typically ranging from several thousand up to a million bosons in size. Implications are briefly discussed.
\end{abstract}


\maketitle

Ultracold bosonic gases under rotation are suitable to probe various phenomena of correlated quantum systems. During the past two decades, rotating Bose-Einstein condensates (BECs) were studied from multiple perspectives, ranging from the occurence of quantized vortices \cite{Rokhsar,Matthews,Madison1,Madison2,Madison3} to vortex lattices and excitations therein \cite{Cornell1,Cornell2,Dalibard1}, and with respect to the analogy to the fractional quantum Hall effect \cite{Regnault,Chang,Dalibard2,Dalibard3}. The literature concerning these topics is extensive and we therefore refer to the reviews in Refs. \cite{Bloch,Cooper,Fetter,Viefers}.

Beside analyzing the ground state of a rotating BEC, low-lying excited states have been of interest because for very low temperatures, they describe the thermodynamic properties of the system. Most studies offering analytical and numerical results for the low-energy spectra were carried out by utilizing the Bogoliubov-de Gennes (BdG) mean-field equation \cite{Bogoliubov,deGennes}, e.g., the decay of the counter-rotating quadrupole mode \cite{Mizushima}, Tkachenko modes in vortex lattices \cite{Simula2, Bigelow2}, the twiston spectrum \cite{Chevy}, or excitations in anharmonic traps \cite{Collin,Ancilotto}. Interestingly, a many-body analysis of the low-energy spectra in rotating BECs is rather rare. Examples are the yrast spectra in a harmonic confinement obtained by exact diagonalization \cite{Reimann1,Reimann2,Viefers2,Reimann3}.

The starting point and motivation of this Letter are different. We consider many-body effects in the low-energy excitation spectrum of a weakly-interacting rotating BEC in a regime where the mean-field theory is supposed to accurately describe the physics. Going beyond previous works, we study bosons in an anharmonic external confinement \cite{Fischer_PRL,Bretin,Yngvason} where one can no longer rely on the validity of the lowest Landau level approximation. The latter is well-suited for rapidly-rotating and slightly-repulsive bosons in a harmonic trap with rotation frequency very close to the trap frequency. 

Rotating the BEC leads to a transfer of high-lying excited states in the laboratory frame to the low-energy part of the spectrum in the rotating  frame. A central role in our anaylsis would be the dependence of the excitation energies and their many-body characteristics on the particle number $N$. It has been shown recently that for a non-rotating repulsive BEC in a trap, the excitation energies in the Hartree limit converge towards the BdG spectrum \cite{Seiringer1}, and similarly for a rotating BEC under certain conditions \cite{Seiringer2}. However, it remains unclear if many-body effects in the excitation spectrum can be observed for mesoscopic and large BECs, typically of the experimentally relevant order of $10^3-10^6$ bosons. Our numerical results present strong physical trends for this regime and show that the answer is positive.

As a main result, we show that once the rotation leads to a quantized vortex in the ground state, substantial many-body effects in the low-energy excitation spectrum occur. These effects grow with the vorticity of the ground state, which means they can be enhanced by stronger rotation. In addition, we demonstrate for a low-lying excited state which is also accessible within mean-field theory, that these effects clearly persist when the number of bosons is increased up to the experimentally relevant regime, despite the BEC being essentially condensed. The present work reports on accurate many-body excitation energies of a two-dimensional BEC obtained by linear-response, and goes well beyond previous investigations of one-dimensional systems \cite{Theisen,Beinke}.

The general form of the Hamiltonian for $N$ interacting bosons in the rotating frame is given by
\begin{equation}\label{Hamiltonian}
 \hat{H}_{\text{rot}}=\hat{H}_{\text{lab}}-\Omega\hat{L}_z, \quad  \hat{H}_{\text{lab}}=\sum_{i=1}^N \hat{h}(\vec{r}_i)+\lambda_0\sum_{i<j}^N \hat{W}(|\vec{r}_i-\vec{r}_j|),
\end{equation}
where the single-particle Hamiltonian is $\hat{h}=-\frac{\Delta}{2}+V$ with the Laplacian $\Delta=\partial^2/\partial\vec{r}^{\,2}$ and the external trapping potential $V$, and the two-body interaction is $\lambda_0 \hat{W}$ with $\lambda_0$ being its strength. The rotation term contains the rotation frequency $\Omega$ and the total angular momentum operator in z-direction, $\hat{L}_z=\sum_{i=1}^N\hat{l}_z(i)$. We work in dimensionless units obtained by dividing $\hat{H}_{\text{rot}}$ by $\frac{\hbar^2}{d^2m}$ where $d$ is a length scale and $m$ the boson mass. A translation to dimensionfull units is given in \cite{params}. 

The Gaussian-shaped repulsion $\lambda_0 \hat{W}(|\vec{r}_i-\vec{r}_j|)=\frac{\lambda_0}{2\pi\sigma^2}\,e^{-|\vec{r}_i-\vec{r}_j|^2/2\sigma^2}$, $\sigma=0.25$ avoids the regularization problems of the zero-ranged contact potential in two dimensions \cite{Doganov} and has been employed in recent works \cite{BeinkeMBTun,Klaiman,Christensson}. The interaction strength $\lambda_0$ for different particle numbers is adjusted such that the mean-field interaction parameter $\Lambda=\lambda_0(N-1)$ is kept constant, i.e., $\lambda_0\sim (N-1)^{-1}$. The trapping potential models a radially-symmetric crater given by $V(r)=C\,e^{-0.5\,(r-R_C)^4}$ for $r=\sqrt{x^2+y^2}\leq R_C$ and $V(r)=C$ for $r> R_C$. The values of the crater height $C$ and the radial size $R_C$ are given in \cite{params}. In contrast to the commonly considered harmonic trapping potential for rotating BECs, this potential ensures that the center-of-mass and relative coordinates do not separate. Furthermore, there is no formation of distinct Landau levels and it is thus required to go beyond the lowest Landau level approximation \cite{SI}.

The standard strategy to compute excited states in a (weakly-interacting) BEC is to first calculate the ground state using the Gross-Pitaevskii (GP) equation \cite{Book1,Book2,Gross,Pitaevskii}. In the rotating frame, it reads \small$\left[\hat{h}+\Lambda\int d\vec{r}^{\,\prime}\hat{W}(|\vec{r}-\vec{r}^{\,\prime}|)|\phi_\text{\tiny{GP}}(\vec{r}^{\,\prime})|^2-\Omega\hat{l}_z\right]\phi_\text{\tiny{GP}}(\vec{r})=\mu\,\phi_\text{\tiny{GP}}(\vec{r})$\normalsize, where $\phi_\text{\tiny{GP}}$ is the ground-state orbital and $\mu$ the chemical potential. Afterwards, one applies linear-response theory atop $\phi_\text{\tiny{GP}}$ which yields the BdG equation,
\begin{equation}\label{BdG_eq}
	\mathcal{L}_\text{BdG} \begin{pmatrix} u^k\\ v^k \end{pmatrix} = \omega_k \begin{pmatrix} u^k\\ v^k \end{pmatrix},
\end{equation}
with the BdG matrix $\mathcal{L}_\text{BdG}$, the correction amplitudes $u_k$ and $v_k$ of the $k$-th excited state to the ground-state orbital, and the excitation energies $\omega_k=E_k-E_0$ relative to the ground-state energy $E_0$. We employ the particle-conserving version of Eq. (\ref{BdG_eq}) \cite{Castin,Gardiner,Castin2,Ruprecht}. It is worth noting that the BdG theory by construction only gives access to excitations where a single boson is excited from the condensed mode.

It is a well-known fact that a linear-response analysis atop the exact ground state gives rise to the exact excitation spectrum \cite{Walecka}. Thus, to go beyond the mean-field approach described above, we increase the accuracy of the ground state by utilizing a many-body ansatz for the wave function, $|\Psi(t)\rangle=\sum_{\vec{n}}C_{\vec{n}}(t)\,|\vec{n};t\rangle$, which is a superposition of permanents $\{|\vec{n};t\rangle\}$ comprised of $M$ single-particle orbitals $\{\phi_j(\vec{r},t):1 \leq j \leq M\}$ and expansion coefficients $\{C_{\vec{n}}(t)\}$ where $\vec{n}=(n_1,\ldots,n_M)^t$ is a vector carrying the individual occupation numbers of the orbitals for a given permanent. Both the orbitals and coefficients are time-adaptive and determined by the Dirac-Frenkel variational principle, yielding the multiconfigurational time-dependent Hartree for bosons (MCTDHB) method \cite{MCTDHB1,MCTDHB2}, see, e.g., \cite{Sakmann2,Bolsinger,Brouzos,Mistakidis,Grond,Fischer_PRA} for applications. 

The ground-state depletion $f$ is defined as the occupation of all but the first natural orbital, $f=\sum_{k>1}^M n_k$. The natural orbitals are the eigenvectors of the one-body reduced density matrix $\bm{\rho}=\{\rho_{ij}\}$ with $\rho_{ij}=\langle\Psi|\hat{b}_i^\dagger \hat{b}_j|\Psi\rangle$ where the annihilation (creation) operator $\hat{b}_i^{(\dagger)}$ removes (adds) a particle from (to) the orbital $\phi_i$. If only the largest occupation $n_1$ is macroscopic, the system is said to be condensed. This is the case in this work since the repulsion between the bosons is very weak. If more than a single occupation is macroscopic, the system is said to be fragmented, and there are recent works dealing with fragmented rotating BECs as well \cite{BeinkeMBTun,Kasevich,DalibardNature,Storm}. 

After computing the many-body ground state, we apply many-body linear-response (LR) theory, termed LR-MCTDHB \cite{LR-MCTDHB1,LR-MCTDHB2}, atop it, also see \cite{Theisen,Beinke}. This leads to an eigenvalue equation of the form
\begin{equation}\label{eigenvalue_eq}
	\mathcal{L}
	  \begin{pmatrix}\bold{u}^k\\\bold{v}^k\\\bold{C}_u^k\\ \bold{C}_v^k \end{pmatrix}
	 =\omega_k \begin{pmatrix}\bold{u}^k\\\bold{v}^k\\\bold{C}_u^k\\ \bold{C}_v^k \end{pmatrix}
\end{equation}
with the $(2M+N_\text{conf})$-dimensional linear-response matrix $\mathcal{L}$ where $N_\text{conf}=\binom{N+M-1}{N}$ is the number of possibilities to distribute $N$ bosons among $M$ orbitals. It consists of four blocks, $\mathcal{L}=\begin{pmatrix} \bold{\mathcal{L}}_{oo} & \bold{\mathcal{L}}_{oc}	\\ \bold{\mathcal{L}}_{co} & \bold{\mathcal{L}}_{cc}	\end{pmatrix}$, accounting for the couplings between the orbitals and coefficients. A detailed derivation of $\mathcal{L}$ and its submatrices is shown in \cite{LR-MCTDHB1,LR-MCTDHB2}. The eigenvector $(\bold{u}^k,\bold{v}^k,\bold{C}_u^k,\bold{C}_v^k)^T$ collects the correction amplitudes to the ground-state orbitals and coefficients, and the eigenvalue $\omega_k=E_k-E_0$ denotes the energy of the $k$-th excited state relative to the ground-state energy $E_0$. We stress that LR-MCTDHB also accounts for excitations where more than a single boson is excited from the condensed mode. It is further important to note that for $M=1$ Eqs. (\ref{BdG_eq}) and (\ref{eigenvalue_eq}) become identical such that the BdG theory is contained in our many-body approach as the simplest limiting case. 

Throughout this Letter, we refer to excitations calculated with  Eq. (\ref{BdG_eq}) as mean-field excitations, whereas to excitations calculated with Eq. (\ref{eigenvalue_eq}) as many-body excitations. Furthermore, it is useful to distinguish between single- and multi-particle excitations where either one or multiple particles are excited from the condensed mode.

Technically, we calculate the lowest few eigenvalues of $\mathcal{L}$ by using the Implicitly Restarted Arnoldi Method \cite{Saad}, a generalization of the Lanczos method for non-hermitian matrices, and its parallel implementation in the ARPACK numerical library \cite{Arpack_HP}. It allows us to treat even large systems with $N=1000$ bosons and $M=3$ orbitals, where already the dimensionality of the coefficient matrix $\bold{\mathcal{L}}_{cc}$ exceeds $10^6$. The numerical results below are converged both with respect to the number of orbitals $M$ and the number of grid points on which the Hamiltonian in Eq. (\ref{Hamiltonian}) is represented \cite{SI}. To calculate the ground states we use the MCTDHB implementation in \cite{package}.

\begin{figure}
\includegraphics[angle=0,width=\columnwidth]{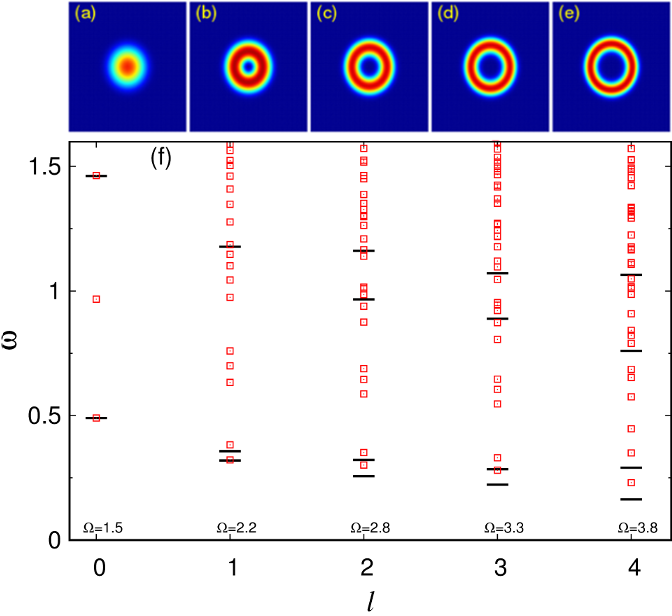}
\caption{(Color online) (a)-(e) Many-body ground-state densities in a rotating BEC of $N=10$ bosons with interaction parameter $\Lambda=0.5$ for vorticities $l=0$ to $l=4$ [panel (a) through (e)]. For $l>0$ the ground state is a single vortex whose size grows with $l$. The mean-field densities are alike (not shown). (f) Corresponding low-energy excitation spectra for the ground states in the upper panels. Thick black lines indicate mean-field results from Eq. (\ref{BdG_eq}), i.e., with $M=1$ (BdG), and red squares denote the many-body results from Eq. (\ref{eigenvalue_eq}) with $M=7$ orbitals (LR-MCTDHB). The many-body spectra show a very rich structure which cannot be accounted for within the mean-field picture. All quantities are dimensionless. See text for details.}
\label{fig:Fig1}
\end{figure}

Fig. \ref{fig:Fig1} shows the ground-state densities and the low-energy excitation spectra of a rotating BEC with $N=10$ bosons and interaction parameter $\Lambda=0.5$ for different rotation frequencies $\Omega$. The many-body energies have been computed utilizing $M=7$ orbitals which ensures numerical convergence for the shown energy range \cite{SI}. The rotation frequencies were chosen such that the underlying ground states have different vorticities, i.e., angular momenta per particle, $l$. The degree of condensation $1-f$ ranges from $9.999$ ($l=0$) to $9.962$ ($l=4$) out of 10 particles, meaning that the BEC is highly condensed and one might expect the mean-field theory to give accurate results. One can see from the densities in Figs. \ref{fig:Fig1}(a)-(e) that the radial symmetry is, of course, preserved under rotation and that the core size of the vortex is growing with vorticity $l$. This is a giant vortex which an anharmonic trap can sustain \cite{Fischer_PRL,Yngvason}, albeit here described at the many-body level. 

With regard to the excitation spectra in Fig.\ref{fig:Fig1}(f), we observe that for $l=0$, i.e., when the ground state is fully condensed, the mean-field and many-body energies of the first two single-particle excitations are equal (first and third state from below). They refer to the cases of taking one boson from the condensed mode to either an orbital with angular momentum $l_z=+1$ or $+2$, respectively. We refer to the former excitation as $(+1)$. How its energy $\omega_{(+1)}$ depends on the vorticity and the number of particles is discussed in detail below. The second excitation from below, only captured at the many-body level, is a two-particle excitation where two bosons occupy the orbital with $l_z=+1$. 

For non-zero ground-state vorticities ($l>0$), the deviations between the BdG and many-body spectra grow substantially. At the many-body level, the increased rotation transfers many more states to the low-energy spectrum than at the mean-field level. Moreover, the differences between BdG and many-body energies of single-particle excitations grow. The inaccuracy of the mean-field energies for single-particle excitations is intriguing since one might expect this simplest kind of excitations to be the least sensitive to many-body effects. In the remaining part of this Letter, we show that this intuition is misleading and that one needs an accurate many-body description even for the lowest single-particle excited states. Therefore, we elaborate on $(+1)$ in more detail.

\begin{figure}
\includegraphics[angle=-90,width=\columnwidth]{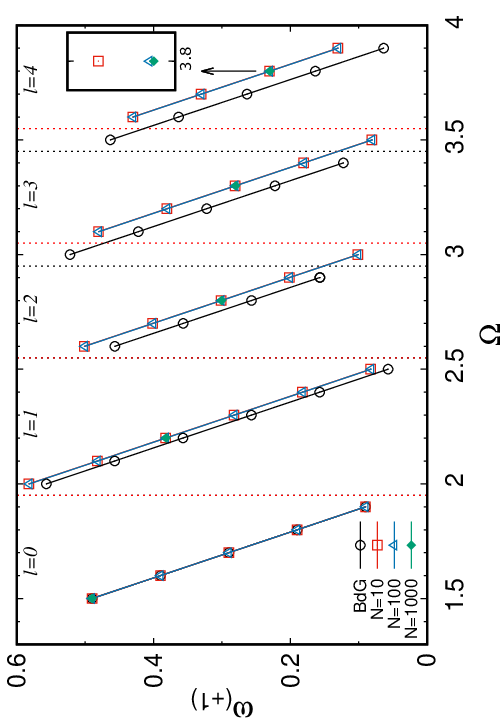}
\caption{(Color online) Enhanced many-body effects by rotation. The excitation energies $\omega_{(+1)}$ are shown as a function of the rotation frequency $\Omega$ for interaction parameter $\Lambda=0.5$. Many-body results are calculated for different particle numbers $N$. Vertical dotted lines indicate the transition from ground-state vorticity $l$ to $l+1$ between two adjacent analyzed rotation frequencies (mean-field in black and many-body in red). It is seen that the many-body effects grow with growing vorticity. Even the assignment of the vorticity to the mean-field and many-body ground states does not match as the vorticity grows. The inset shows a magnified view for $\Omega=3.8$. All quantities are dimensionless. See text for details.}
\label{fig:Fig2}
\end{figure}

Fig. \ref{fig:Fig2} shows the excitation energy $\omega_{(+1)}$ for a broader range of rotation frequencies $\Omega$ and compares mean-field and many-body energies for $N=10, \, 100$, and $1000$ bosons. Up to $\Omega=1.9$ ($l=0$), the mean-field and many-body results coincide, meaning that rotating the BEC with $\Omega\leq 1.9$ does not lead to visible many-body corrections to the excitation energy $\omega_{(+1)}$.

Once the ground state of the BEC becomes a vortex ($\Omega\geq 2.0$), the mean-field and many-body results start to deviate. The energetic distance between them grows with $l$. This can be rationalized with the geometry of a vortex. Since it has the shape of a ring with finite radial width, a comparison to a one-dimensional system of interacting bosons on a finite ring is instrumental. For such a system, it has been demonstrated recently that the overlap between the mean-field and the exact ground state decreases and that the depletion grows with the size of the ring \cite{Cederbaum2017}. This implies here that the quality of the BdG results decreases with growing vortex size and thus with growing vorticity. Nonetheless, we point out again that even for $N=10$ and $l=4$, $f$ is only $0.038$. Thus the BEC is highly condensed and one might \textit{a priori} expect the BdG equation to yield accurate results for the lowest excitation energies. Instead, for $\Omega=3.9$, the difference between the mean-field and many-body energies becomes approximately as large as the mean-field energy itself. 

The inset in Fig. \ref{fig:Fig2} magnifies $\omega_{(+1)}$ with respect to the particle number $N$ for $\Omega=3.8$. One can see that the excitation energy decreases as $N$ grows, but increasing $N$ from $100$ to $1000$ lowers $\omega_{(+1)}$ only marginally and one is rather far from the mean-field result. 

\begin{figure}
\includegraphics[angle=-90,width=\columnwidth]{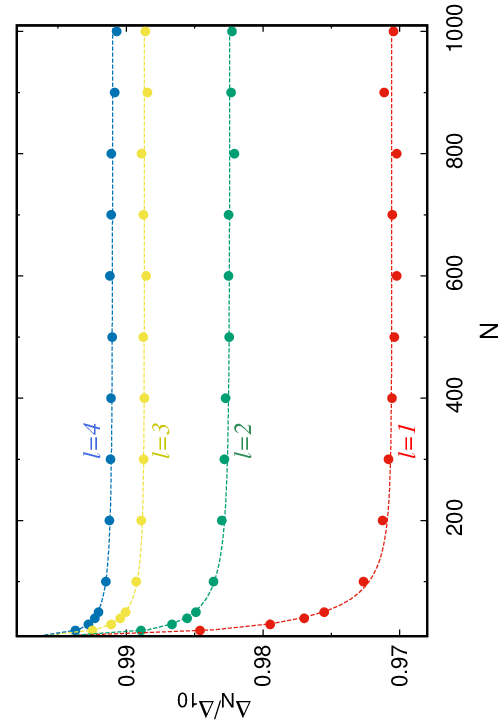}
\caption{(Color online) Gap size $\Delta_N$ for particle numbers from $N=20$ to $N=1000$. Results are given relative to $\Delta_{10}$, the dashed lines indicate exponential-fit curves. For $l=0$ $\Delta_N$ is essentially zero (not shown). The gap varies with $N$ weaker as the vorticity $l$ grows. Small oscillations in the tails of the curves $\Delta_N/\Delta_{10}$ are due to the numerical accuracy of order $\mathcal{O}(10^{-5})$. All quantities are dimensionless. See text for details.}
\label{fig:Fig3}
\end{figure}

Fig. \ref{fig:Fig3} shows the impact of $N$ on the energy gap $\Delta_N=\omega_{(+1)}^N-\omega_{(+1)}^{\text{\tiny{BdG}}}$, where $\omega_{(+1)}^N$ is the many-body and $\omega_{(+1)}^{\text{\tiny{BdG}}}$ the mean-field energy of $(+1)$. Using $M=3$ orbitals ensures numerical convergence for the obtained results of $\omega_{(+1)}^N$ \cite{SI}. The gap size is shown relative to $\Delta_{10}$, i.e., for $N=10$ bosons. For all values of $l$, $\Delta_N$ decreases until $N \approx 200$ and then apparently slowly saturates. Moreover, the gap varies with $N$ weaker as the vorticity $l$ grows. Even for $l=1$, the size of $\Delta_N$ remains around $97\%$ of $\Delta_{10}$ for $N=1000$, and it remains even larger for higher values of $l$. We stress again that the BEC is highly condensed for all chosen rotation frequencies. As an example, the degree of condensation for $N=1000$ and $l=4$ is $999.97$, i.e., only $f\approx 0.03$ bosons (or 0.003$\%$) are outside the condensed mode. According to the results in Fig. \ref{fig:Fig3} there is basically no evidence that $\Delta_N$ would show a sharp descent when $N$ is increased by two or three orders of magnitude beyond the particle numbers considered in this work. Naturally, the question arises whether the asymptotic behavior discussed in Refs. \cite{Seiringer1,Seiringer2}, namely that the BdG spectrum yields the exact excitation energies for both trapped non-rotating and rotating but symmetry-broken BECs, is the same for a rotating BEC where the radial symmetry is preserved. Although our results do not answer this question in general, they at least indicate that in such a case one would need an unconceivable amount of bosons to come close to the BdG energies, certainly more than $10^4-10^6$ bosons which are typically used in experiments.

To summarize, we have shown that rotating a weakly-interacting BEC leads to a strong enhancement of many-body effects in the low-energy excitation spectrum, even if the degree of condensation is very high. Beside the fact that the amount of multi-particle excited states increases strongly when the ground state is a vortex, the differences between the mean-field and accurate many-body excitation energies grow with growing vorticity and can become of the order of the mean-field excitation energies themselves, even for the very lowest single-particle excited states. Moreover, these differences clearly persist for larger particle numbers, showing that an accurate many-body theory for the low-energy excitation spectrum is necessary, not only for a small amount of bosons. Such many-body effects in the excitation spectrum would be interesting to search for also in the dynamics of essentially-condensed rotating BECs.

\begin{acknowledgements}
\textit{Acknowledgements.} We thank Alexej I. Streltsov and Shachar Klaiman for many discussions. Computation time on the Cray XC40 cluster Hazel Hen at the High Performance Computing Center Stuttgart (HLRS) and the BwForCluster is acknowledged. RB acknowledges financial support by the IMPRS-QD (International Max Plack Research School for Quantum Dynamics), the Landesgraduiertenf{\"o}rderung Baden-W{\"u}rttemberg, and the Minerva Foundation. OEA acknowledges funding by the Israel Science Foundation (Grant No. 600/15).
\end{acknowledgements}



\begin{thebibliography}{99}
\bibitem{params}	We use $R_C=3.0$ and $C=200.0$, meaning that $C$ is higher than the energy per particle and the BEC is bound by the crater. The trap is represented on a grid of $64\times 64$ grid points on a two-dimensional box with range $[-4,4)\times[-4,4)$ \cite{SI}. To convert to real units, we suggest a condensate of $^{87}$Rb atoms and a length scale of $d=2\,\mu$m which yields $182.7$ Hz as a unit of energy, such that the crater height is $C=36.5$ kHz. With $R_C=6\,\mu$m and an aspect ratio of $\frac{a_z}{R_C}=\frac{1}{200}$ between $R_C$ and the transverse confinement $a_z$ \cite{BeinkeMBTun}, the scattering length $a_s=\frac{\Lambda\,a_z}{(N-1)2\sqrt{2\pi}}$ \cite{params2} for $N=10$ is given by $a_s=0.67$ nm, and correspondingly lower values for higher boson numbers.
\end{thebibliography}
\end{document}